\documentclass{elsart}
\usepackage{url,amsfonts}
\journal{Theoretical Computer Science}

\newcommand{\A}{\mathcal{A}}
\newcommand{\E}{\mathcal{E}}
\newcommand{\group}{\mathcal{G}}
\newcommand{\C}{\mathcal{C}}
\newcommand{\Se}{\mathcal{S}}
\newcommand{\Z}{\mathbb{Z}}

\newcommand{\hash}{\mathcal{H}}

\begin{document}
\begin{frontmatter}

\title{Security Analysis of a Password-Based Authentication
Protocol Proposed to IEEE 1363}
\author[zhu]{Zhu Zhao},
\ead{zhaozhu@hxu.edu.cn}
\author[zhongqi]{Zhongqi Dong},
\ead{dongzha03@st.lzu.edu.cn}
\author[yongge]{Yongge Wang} %\corauthref{cor}}
%\corauth[cor]{Corresponding author.}

\address[zhu]{Hexi University, ZhangYe City, GanSu Province, P.R.China}
\address[zhongqi]{Lanzhou University, Lanzhou City, GanSu Province, P.R.China}
\address[yongge]{SIS Department, UNC Charlotte, Charlotte, NC, USA}
\ead{yonwang@uncc.edu}

\begin{abstract}
In recent years, several protocols for password-based authenticated key
exchange have been proposed. These protocols aim to be secure even though
the sample space of passwords may be small enough to be enumerated by
an off-line adversary. In Eurocrypt 2000, Bellare, Pointcheval and 
Rogaway (BPR) presented a model and security definition for authenticated
key exchange. They claimed that in the ideal-cipher model (random oracles), 
the two-flow protocol at the core of Encrypted Key Exchange (EKE) is 
secure.  Bellare and Rogaway suggested several instantiations of the ideal 
cipher in their proposal to the IEEE P1363.2 working group. 
Since then there has been an increased interest
in proving the security of password-based protocols in the ideal-cipher model.
For example, Bresson, Chevassut, and Pointcheval have recently showed
that the OEKE protocol is secure in the ideal cipher model.
In this paper, we present examples of \emph{real} (NOT ideal) ciphers 
(including naive implementations of the instantiations proposed to 
IEEE P1363.2) that would result in broken instantiations of the 
idealized AuthA protocol and OEKE protocol. Our result shows that 
the AuthA protocol can be instantiated in an insecure way, and that 
there are
no well defined (let alone rigorous) ways to distinguish between secure and
insecure instantiations. Thus, without a rigorous metric for ideal-ciphers,
the value of provable security in ideal cipher model is limited.
\end{abstract}

\end{frontmatter}

\section{Introduction}
Numerous cryptographic protocols rely on passwords selected by users
(people) for strong authentication.  Since the users find it inconvenient
to remember long passwords, they typically select short easily-rememberable 
passwords. In these cases, the sample space of passwords may be small enough 
to be enumerated by an adversary thereby making the protocols vulnerable to
a \emph{dictionary} attack. It is desirable then to design password-based
protocols that resist off-line dictionary attacks. 

The password-based protocol problem was first studied by 
Gong, Lomas, Needham, and Saltzer \cite{gong} who used public-key
encryption to guard against off-line password-guessing attacks. 
In another very influential work \cite{eke}, Bellovin and Merritt 
introduced Encrypted Key Exchange (EKE), which became the basis
for many of the subsequent works in this area. These protocols
include SPEKE \cite{speke} and SRP \cite{srp,srp6}. 
Other papers addressing the above protocol problem can be found in
\cite{pak,besw,hk,lucks}.
Bellare, Pointcheval, and Rogaway \cite{bpr} defined
a model for the password-based protocol problem and claimed that their model
is rich enough to deal with password guessing, forward secrecy,
server compromise, and loss of session keys. Then they claimed
that in the ideal-cipher model (random oracles), the two-flow protocol
at the core of Encrypted Key-Exchange (EKE) is secure. In addition,
Bellare and Rogaway \cite{autha} suggested several instantiations 
(AuthA) of the ideal-cipher
in their proposal to the IEEE P1363.2 Working Group. 
Recently, Bresson, Chevassut, and Pointcheval \cite{ccs03}
proposed a simplified version of AuthA, which is called OEKE, and showed
that OEKE achieves provable security
against dictionary attacks in both the random oracle and ideal-cipher
models under the computational Diffie-Hellman intractability assumption.

The ideal-cipher model was introduced by 
Bellare, Pointcheval, and Rogaway \cite{bpr} as follows.
Fix finite sets of strings ${\mathcal G}$ and  ${\mathcal C}$
where $|{\mathcal G}|= |{\mathcal C}|$. In the ideal-cipher model,
choosing a random function $h$ from $\Omega$ amounts to
giving the protocol (and the adversary) a perfect way to 
encipher strings in  ${\mathcal G}$: namely, for $K\in\{0,1\}^*$, we 
set ${\mathcal E}_K: {\mathcal G}\rightarrow {\mathcal C}$ to be a random
bijective function, and we let 
${\mathcal D}_K: \{0,1\}^*\rightarrow {\mathcal G}$
defined by ${\mathcal D}_K(y)$ be the value $x$ such that 
${\mathcal E}_K(x)=y$, if $y\in {\mathcal C}$, and undefined otherwise.

This paper studies the security issues with 
practical realization of the ideal cipher model by
Bellare, Pointcheval, and Rogaway \cite{bpr}.
We show that for several instantiations of the 
ideal-cipher (including naive implementations of instantiations suggested in
\cite{1363a}), the instantiated Bellare and Rogaway's protocol
(AuthA) is not secure against off-line dictionary attacks. Our results show
that realizing the ideal-cipher of  Bellare, Pointcheval, and Rogaway
can be tricky. In particular, our results point out the weakness
in the ideal-cipher methodology of Bellare, Pointcheval, and Rogaway.
That is, without a robust measuring method for deciding whether a
given cipher is a ``good realization'' of the ideal-cipher, 
ideal-cipher model analysis \cite{bpr,ccs03} of a password-based protocol can 
be of limited value. Indeed, there is no well defined (let alone rigorous) 
way in \cite{bpr} to distinguish between secure and insecure instantiations 
of an ideal-cipher. 
Note that Black and Rogaway \cite{black} have done some initial research
on the potential implementations of ideal-ciphers with
arbitrary finite domains. However, it is still far from a complete solution.

One of the main applications of password-based protocols is in the 
environment of wireless and other more constrained devices (e.g., 
secure downloading of private credentials: SACRED \cite{sacred}). 
Elliptic Curve Cryptography (ECC) has been extensively used 
in these constrained devices. However, most of the suggested 
password-based protocols are described in the group (or subgroup of) 
$\group=\Z^*_p$, and are  either non-friendly or non-secure for 
ECC-based groups. For example, SRP \cite{srp,srp6} is based on a 
field and used both field operations of addition and multiplication, but 
ECC groups only have one group operation.
Several ECC-based SRP protocols have been introduced in 
Lee and Lee \cite{ecsrp}. We will show that one of these protocols 
is completely insecure. We will also discuss the security issues related
ECC-based SRP protocols. 
As an example, we will also present a variant SRP5 of the 
original SRP protocol.

The organization of the paper is as follows: In Section~\ref{security}
we informally address the security problems of password-based protocols.
We mount attacks on several instantiations of 
Bellare, Pointcheval, and Rogaway's AuthA protocol and on instantiations of
Bresson, Chevassut, and Pointcheval's OEKE protocol
in Sections~\ref{survey} and \ref{oekesec} respectively. 
In Section~\ref{sec-srp} we briefly discuss instantiations of OEKE 
and the  SRP protocol. We draw our conclusions in Section~\ref{conclusions}.

\section{Security of password authentication}
\label{security}

Halevi and Krawczyk \cite[Sections 2.2-2.3]{hk} introduced
a notion of security for password authentication. 
They provide a list of basic attacks that a password-based protocol 
needs to guard against. In the following, we provide the 
list of attacks. An ideal password protocol should be 
secure against these attacks and we will follow these
criteria when we discuss the security of password protocols.
\begin{itemize}
\item {\bf Eavesdropping}. The attacker may observe the communications 
channel.
\item {\bf Replay}. The attacker records messages she has observed
and re-sends them at a later time.
\item {\bf Man-in-the-middle}. The attacker intercepts the 
messages sent between the parties  $\C$ and $\Se$ and replaces these with her
own messages. She plays the role of the client in the messages
which it sends to the server, and at the same time plays the role of the 
server in the messages that she sends to the client. A special
man-in-the-middle attack is the \emph{small subgroup attack} 
\cite{ll,subgroup,vw}. We illustrate this kind of attack by a small
example. Let $g$ be a generator of the group $\group$ of 
order $n=qt$ for some small $t>1$. In a standard
Diffie-Hellman key exchange protocol, the client $\C$ chooses a 
random $x$ and sends $g^x$ to the server 
$\Se$, then $\Se$ chooses a random $y$ and sends $g^y$ to $\C$.
The shared key between $\C$ and $\Se$ is $g^{xy}$.
Now assume that the attacker $\A$ intercepts $\C$'s message $g^x$,
replaces it with $g^{xq}$, and sends it to $\Se$. $\A$ also intercepts
$\Se$'s message $g^y$, replaces it with  $g^{yq}$, and sends it to $\C$.
In the end, both $\C$ and $\Se$ compute the shared key $g^{qxy}$.
Since $g^{qxy}$ lies in the subgroup of order $t$ of the group
generated by $g^q$, it takes on one of only $t$ possible values.
$\A$ can easily recover this  $g^{qxy}$ by an exhaustive search.
\item {\bf Impersonation}. The attacker impersonates the client
or the server to get some useful information.
\item {\bf Password-guessing}. The attacker is assumed to have 
access to a relatively small dictionary of words that likely includes the 
secret password $\alpha$. In an \emph{off-line attack}, the attacker 
records past communications and searches for a word in the dictionary that
is consistent with the recorded communications. In an \emph{on-line attack}, 
the attacker repeatedly picks a password from the dictionary and attempts 
to impersonate $\C$ or $\Se$. If the impersonation fails, the attacker
removes this password from the dictionary and tries again, using 
a different password. 
\item {\bf Partition attack}. The attacker records past communications,
then goes over the dictionary and deletes those words that are not  
consistent with the recorded communications from the dictionary. After 
several tries, the attacker's dictionary could become very small.
\end{itemize}
We now informally sketch the definition of security in \cite{hk} for 
a password-based protocol.  The attacker $\A$ is allowed to watch 
regular runs of the protocol between the client $\C$ and the server
$\Se$, and can also actively communicate with $\C$ and $\Se$
in replay, impersonation, and man-in-the-middle attacks.
A protocol is said to be \emph{secure} in the presence of such an 
attacker if (i) whenever the server $\Se$ accepts an authentication
session with $\C$, it is the case that $\C$ did indeed participate in 
the authentication session; and (ii) $\C$ accepts an authentication
session with $\Se$, it is the case that $\Se$ did indeed participate in 
the authentication session.

\section{Security issues with practical 
realizations of the ideal cipher model: 
on Bellare and Rogaway's AuthA}
\label{survey}
In the remainder of this paper, we will use the following notations:
By $\group=\langle g\rangle$, we denote a cyclic group generated by $g$,
and by $ord(g)$, we denote the order of $g$. 
For a symmetric
encryption scheme $\E$ and a key $\pi$, $\E_\pi(x)$ denotes
the ciphertext of $x$. We also assume that the client $\C$ 
holds a password $\alpha$ and the server $\Se$ holds a key 
$\beta$ which is a known function of $\alpha$.
In a protocol for a symmetric model,
the client and the server share the same password, that is, 
$\beta=\alpha$. In this paper, we will abuse our notation
by letting $\C$ and $\Se$ also denote corresponding parties' identification
strings.
In a protocol for an asymmetric model, $\beta$ will typically
be chosen so that it is hard to compute $\alpha$ from $\C$,
$\Se$, and $\beta$. The password $\alpha$ might be a poor one. 
Probably the user selects some short easily-rememberable $\alpha$
and then installed $\beta$ at the server.
In the protocols, $\hash$ is used to denote a secure hash function.
We will also abuse our notation by using $\C$ (respectively, $\Se$)
to denote the identity number of the client (respectively, the server).

\subsection{The AuthA protocol}
Bellare, Pointcheval, and Rogaway \cite{bpr} defined
a model for the password-based protocol problem and showed that their model
is rich enough to deal with password guessing, forward secrecy,
server compromise, and loss of session keys. Then they proved
that in the ideal-cipher model (random oracles), the two-flow protocol
at the core of Encrypted Key Exchange (EKE)
is secure. In addition, Bellare and Rogaway \cite{autha}
suggested several instantiations of the ideal-cipher
in their proposal to IEEE P1363.2 working group.  In the protocol, 
the server $\Se$ stores the value $\langle \C, \beta\rangle$ 
for each client $\C$ where $\beta=g^\alpha$. The protocol proceeds as follows:
\begin{enumerate}
\item $\C$ chooses a random $x\in [1, ord(g)-1]$, computes
$g^x$, encrypts it with $\beta$, 
and sends the ciphertext $\E_{\beta}(g^x)$ to the Server $\Se$.
\item $\Se$ chooses a random $y\in [1, ord(g)-1]$, computes
$g^y$, encrypts it with $\beta$, 
and sends the ciphertext $\E_{\beta}(g^y)$ to $\C$.
\item AuthA authentication 
steps. Let $K=\hash(\C||\Se||g^x||g^y||g^{xy})$.
Then there are three authentication methods for AuthA:
\begin{enumerate} 
\item The server authenticates himself by
sending $\hash(K||2)$ to $\C$.
\item The client authenticates himself by sending 
$\hash(K||g^{\alpha y})$ to $\Se$.
\item Both server and client achieve mutual authentication by
sending both of the messages in the above two steps
\end{enumerate}
\end{enumerate}

The authors of \cite{bpr} claimed that if the encryption 
function $\E$ is given by an ideal-cipher (random
oracle), then the first-two-step sub-protocol (of AuthA) 
at the core of EKE is provably secure in their model. 
In the following sections, 
we present examples of {\em real} (NOT ideal) ciphers
(including two naive implementations of the three instantiations proposed to 
IEEE P1363.2) that would result in broken instantiations of the 
idealized AuthA protocol.
Indeed, in \cite{bpr}, the authors warn that ``incorrect 
instantiation of the encryption primitive, including instances
which are quite acceptable in other contexts, can easily destroy
the protocol's security''. Our examples confirm this argument.

\subsection{Instantiation 
$\E_{\beta}(X)=X\cdot g^{\hash(\beta)}$}
\label{myinst}
Assume that $\hash$ is a random oracle. 
Bellare and Rogaway \cite{autha} suggested the
instantiation $\E_{\beta}(X)=X\cdot \hash(\beta)$ of 
the ideal-cipher. Obviously, this is far from an ideal cipher.
However, this misleading instantiation will give one 
impression that $\E_{\beta}(X)=X\cdot g^{\hash(\beta)}$ could also be 
a ``reasonable'' instantiation of the ideal-cipher.
Indeed, one may wonder, if $\E_{\beta}(X)=X\cdot \hash(\beta)$
is an ideal cipher, why $\E_{\beta}(X)=X\cdot g^{\hash(\beta)}$ is not?
In the following, we will describe our
attack on the two-step protocol with this instantiation 
$\E_{\beta}(X)=X\cdot g^{\hash(\beta)}$.

No matter whether there is an authentication step (as in AuthA) or not,
our attack works for the two-step protocol. 
If there is an authentication step, then the adversary $\A$
will launch impersonation attacks and use the authentication 
messages to verify whether the guessed password is a correct one.
Without loss of generality, we assume that the server sends 
the first authentication message if any authentication message 
is ever sent between $\C$ and $\Se$ (if the first authentication
message is sent from client to server, then the following
attack works when the adversary impersonate the server). 
If there is no authentication
step, then the adversary could not check whether a guessed
password is a correct one. However, in practice, the established session
key will be used either
to encrypt the actual data for the application protocol or
to encrypt client's private credential (e.g., client's private key).
In either case, the adversary $\A$ can verify whether the guessed
password is a correct one by checking the redundancy in these encrypted data. 
Specifically, consider the following scenario.
$\A$ impersonates the client, chooses a random $z$, 
and sends $g^z$ to the server. The server $\Se$ chooses
a random $y$, sends $g^{y+ \hash(\beta)}$ to $\A$, and computes
the shared key 
$K=\hash(\C||\Se||g^{z-\hash(\beta)}||g^y||g^{(z-\hash(\beta))y})$. 
$\A$ distinguishes the following three cases:
\begin{enumerate}
\item\label{firstcase} This is an AuthA protocol 
and $\Se$ sends $\hash(K||2)$ to 
$\A$ for authentication.
For each guessed $\beta'$, $\A$ computes 
$$K'=\hash(\C||\Se||g^{z-\hash(\beta')}||g^{y+\hash(\beta)-\hash(\beta')}||
g^{(y+\hash(\beta)-\hash(\beta'))(z-\hash(\beta'))}).$$
Note that if $\beta'=\beta$, then $K'=K$ and
$\hash(K||2)= \hash(K'||2)$. Thus $\A$ can decide 
whether $\beta'$ is the correct password.
\item $\Se$ sends $\E_K(m)$ to $\A$, where $m$ is some
application data and has sufficient redundancy. For each
guessed $\beta'$, $\A$ computes $K'$ as in the above item 
\ref{firstcase} and decrypts
$\E_K(m)$ as $m'=\E^{-1}_{K'}(\E_K(m))$. If $\beta'=\beta$, then $K'=K$ and
$m'=m$. Thus by checking the redundancy in $m'$, $\A$ can decide whether 
she has guessed the password correctly.
\item $\Se$ sends $\E_K(\pi)$ to $\A$, where $\pi$ is $\C$'s private key
encrypted with $\C$'s password $\alpha$. Similarly, for each
guessed $\alpha'$, $\A$ first computes $\beta'$, then computes 
$K'$ as in the above item \ref{firstcase} and decrypts $\E_K(\pi)$ as 
$\pi'=\E^{-1}_{K'}(\E_K(\pi))$. If $\beta'=\beta$, then
$K'=K$ and  $\pi'=\pi$. $\A$ further decrypts $\pi'$ with $\alpha'$ to see 
whether the decrypted value is the private key of $\C$. 
Since $\A$ knows $\C$'s public key, she can easily verify this fact. 
Thus, $\A$ can decide whether she has guessed the password correctly.
\end{enumerate}

The above attack demonstrates the inherent weakness
in the ``ideal-cipher model methodology'' by Bellare, Pointcheval, 
and Rogaway \cite{bpr}. That is, without a robust measuring 
method for deciding whether a given cipher is a ``good realization'',
ideal-cipher model analysis of a password-based protocol can be of 
limited value. Indeed, there is no well defined (let alone 
rigorous) way in \cite{bpr} to distinguish between secure and 
insecure instantiations of an ideal-cipher.

\subsection{Instantiation  $\E_{\beta}(X)=X\cdot \hash(\beta)$}
\label{aaa}
The first ideal-cipher instantiation for AuthA
in \cite{autha} is:  $\E_{\beta}(X)=X\cdot \hash(\beta)$.
The authors suggested that the group $\group=\langle g\rangle$ could 
be a group on which the Diffie-Hellman problem is hard:
\begin{quote}
...This group could be $\group = \Z^*_p$, or it could be a prime-order
subgroup of this group, or it could be an elliptic curve 
group...(from \cite{bpr})
\end{quote}
After the introduction of the instantiation function, the authors \cite{autha}
commented that ``you apply the mask generation function
$\hash$ to $\beta$, interpret the result as a group element,
and multiply by the plaintext''. However, for most implementations,
one may ignore this comment and just multiply the hash
result with the plaintext. Naively, one can also interpret the hash result
$\hash(\beta)$ as a group element $g^{\hash(\beta)}$. Then our attacks in
Section \ref{myinst} show that this instantiation is not secure.
Indeed, from the ideal-cipher assumption, it is not clear that
one needs to interpret the hash result as a group element other than
$g^{\hash(\beta)}$. One may feel that  
both $X\cdot \hash(\beta)$ and $X\cdot g^{\hash(\beta)}$
can be regarded as acceptable instantiations of the ideal cipher
over $\Z^*_p$ (why not?).
In the following, we mount an off-line dictionary attack on this 
instantiation without interpreting the result as a group element.

Our attack in Section \ref{myinst} 
does not work for AuthA with this instantiation. However, we can 
show that this instantiation will leak some information of 
the password $\alpha$ if the group is a subgroup of  $\Z^*_p$
or an  elliptic curve group.
As an example, we illustrate the information leakage of AuthA 
with a subgroup of $\Z^*_p$.
Assume that $p=tq+1$ with $\gcd(t,q)=1$. 
In practice, generally one chooses $p=2q+1$ for some large 
prime $q$ (see, e.g., \cite{mov}), and $ord(g)=q$.

In the attack, the eavesdropper $\A$ intercepts the message
$g^x\cdot\hash(\beta)$, computes $(\hash(\beta))^q
=(g^x\cdot\hash(\beta))^q$. For each guessed
$\beta'$, $\A$ checks whether $(\hash(\beta'))^q=(\hash(\beta))^q$.
If the equation does not hold, then $\A$ deletes $\beta'$ from
her dictionary.
Since $\hash$ is a random oracle, the value
$(\hash(x))^q$ is uniformly distributed over
the set $\{g_1^q, g_1^{2q},\ldots, g_1^{tq}\}$ when $x$ is chosen 
at random, where $g_1$ is a generator of $\Z_p^*$. That is, 
$\Z_p^*=\langle g_1\rangle$.
Thus, $\log t$ bits information of the password is leaked for each
communication between the client and the server with different 
Diffie-Hellman parameters. Thus, 
after $\lceil\frac{|\alpha|}{\log t}\rceil$ observations of
communications between the client and the server with different
Diffie-Hellman parameters, the adversary will recover the 
password with high probability.

Despite the above attacks, we feel that 
AuthA could be securely instantiated by the cipher:
$\E_{\beta}(X)=X\cdot \imath(\hash(\beta))$, where 
$\hash$ is a secure hash function 
and where $\imath$ maps a random string to a group 
element of order $ord(g)$ by ``increasing'' the random string
one by one until reaching a group element with the above given
property. This instantiation should
work both for ECC based groups and for subgroups of
$\Z^*_p$. But we would like to warn that we have not proved with 
reasonable assumptions that this is a secure instantiation of 
the ideal cipher. Of course, it has been proven \cite{bpr} that if 
$\E_{\beta}(X)=X\cdot \imath(\hash(\beta))$
is an ideal cipher then the above instantiation
is provable secure against off-line dictionary attacks. But 
we have no mechanisms to measure whether the above cipher is an ideal cipher.

\subsection{Instantiation  
$\E_{\beta}(X)=(r, X\cdot \hash(r||\beta))$}
The second ideal-cipher instantiation for AuthA
in \cite{autha} is:  $\E_{\beta}(X)=(r, X\cdot \hash(r||\beta))$,
where $r$ is independently chosen at random for each session.
After the introduction of this instantiation, the authors \cite{autha}
did not mention that the hash result $\hash(r||\beta)$ should 
be interpreted as a group element before applying the multiplication.
However we assume that the authors have this in mind when they introduce
this instantiation.
But this again shows that a naive implementation may
multiply the hashing result with $X$ directly without
interpreting it as a group element since 
$\E_{\beta}(X)=(r, X\cdot \hash(r||\beta))$
could be regarded as an acceptable ideal cipher. Indeed,
the ideal cipher model does not address this tiny difference between the 
two implementations: interpreting the hashing result as a group element 
and not interpreting the hashing result as a group element. 

Indeed this instantiation without interpreting
the hashing result as a group element is completely insecure 
against partition attacks if the underlying group is a subgroup of  $\Z^*_p$
or an  elliptic curve group.
The attack in Section \ref{aaa} can be used to show that 
for each randomly chosen $r$, 
$\log t$ bits information of the password $\alpha$ is leaked.
Thus after recording several communications with different
$r$, the adversary can recover $\alpha$.

\subsection{Instantiation  $\E_{\beta}(X)$ by a cipher}
\label{314}
The third ideal-cipher instantiation for AuthA
in \cite{autha} is simply a cipher, e.g., 
$\E_{\beta}(X)=\mathrm{AES}_\beta(X)$.
AuthA with this instantiation is not secure against partition
attacks if the underlying group is a subgroup of  $\Z^*_p$
or an  elliptic curve group. The insecurity of this instantiation
has been observed by several authors, see, e.g., \cite{patel,boyd}.

Firstly we assume that the underlying group $\group$ is a subgroup of 
$\Z_p^*$. The eavesdropper $\A$ tries to decrypt 
$\E_{\beta}(g^x)$ and $\E_{\beta}(g^y)$
with different guessed $\beta'$ ($=g^{\alpha'}$).
If either of the decrypted
value $\E^{-1}_{\beta'}(\E_{\beta}(g^x))$ or 
$\E^{-1}_{\beta'}(\E_{\beta}(g^y))$
is not an element of $\group$, then $\A$ knows that
$\alpha'$ is not the correct password.
Since  $\E_{\beta}(X)$ is an ideal cipher,
only with probability $\left(||\group||/2^{|p|}\right)^2$
both $\E^{-1}_{\beta'}(\E_{\beta}(g^x))$ and
$\E^{-1}_{\beta'}(\E_{\beta}(g^y))$
are elements of $\group$, where $|p|$ and $||\group||$ denote
the length of $p$ in binary representation and the cardinality of
$\group$ respectively.
Thus for each execution of the protocol, $2\log (2^{|p|}/||\group||)$ bits
information of the password $\alpha$ is leaked.
After recording several executions of the protocol,  $\A$ recovers
the password.

Secondly assume that the underlying group $\group$ is an
elliptic curve group. For an elliptic curve group 
$E_{a,b}(F^*_p)=\langle g\rangle$ 
over the field $F^*_p$, the element $(x,y)\in \langle g\rangle$ is denoted by
its $x$ and $y$ coordinates. For a random
chosen $x\in F^*_p$, the probability that there
exists a $y\in F^*_p$ such that $(x,y)$ is a point on the
curve is $1/2$. Thus AuthA over elliptic curve groups 
with this instantiation is not secure against partition attacks.

\section{Security issues with ideal ciphers in one encryption 
key exchange OEKE}
\label{oekesec}
Recently, Bresson, Chevassut, and Pointcheval \cite{ccs03}
formally modeled the AuthA protocol by the One-Encryption-Key-Exchange
(OEKE): only one flow is encrypted (using either a symmetric-encryption
primitive or a multiplicative function as the product of a Diffie-Hellman
value with a hash of the password). The authors pointed out that 
the advantage of OEKE over the classical EKE, wherein the two
Diffie-Hellman values are encrypted, is its easiness of integration.
For example, in Transport Layer Security (TLS) protocol with password-based
key-exchange cipher suits \cite{sbew,taylor}.

OEKE is similar to AuthA except that the first message is not encrypted.
In particular, the protocol proceeds as follows:
\begin{enumerate}
\item $\C$ chooses a random $x\in [1, ord(g)-1]$, computes
$g^x$ and sends $g^x$ to the Server $\Se$.
\item $\Se$ chooses a random $y\in [1, ord(g)-1]$, computes
$g^y$, encrypts it with $\beta$, 
and sends the ciphertext $\E_{\beta}(g^y)$ to $\C$.
\item $\C$ computes $\mbox{Auth}=\hash_1(\C||\Se||g^x||g^y||g^{xy})$,
and sends  $\mbox{Auth}$ to $\Se$.  $\C$ also computes 
session key $K=\hash_0(\C||\Se||g^x||g^y||g^{xy})$.
\item $\Se$ verifies that the value  $\mbox{Auth}$ is correct and computes
the session key similarly.
\end{enumerate}
Where $\hash_0$ and $\hash_1$ are two independent random oracles.
The authors \cite{ccs03} show that the protocol OEKE achieves 
provable security against dictionary attacks in both the random 
oracle and ideal-cipher models under the computational Diffie-Hellman 
intractability assumption.
The authors \cite{ccs03} also observed that a simple block-cipher
could not be used for the instantiation of the ideal-cipher due
to the partition attacks. 

The authors recommended two instantiations of the ideal cipher.
In the first method which is essentially from \cite{bbdp}, 
one encrypts the element, and re-encrypts the result, until one finally
falls in the group $\group$. The second instantiation is the 
cipher  $\E_{\beta}(X)=X\cdot \hash(\beta)$ that we have discussed
in Section \ref{aaa}. That is (see \cite{ccs03}), 
``to instantiate the encryption
primitive as the product of a Diffie-Hellman value with a hash
of the password, as suggested in \cite{autha}''. Obviously, if one 
dose not interpret the hashing output of password as a group element before
applying the multiplication,
then our attacks in Section \ref{aaa} work for OEKE also.
Thus we have the same concern for OEKE: the ideal cipher model does
not directly address the issues of interpreting the hashing output
as group elements. From the ideal cipher model viewpoints,
the two instantiations (one with interpretation of group elements
and one without  interpretation of group elements) have no essential 
difference. However, one instantiation results in broken protocol.
This observation strengthens our viewpoint: without a rigorous way 
to distinguish between secure and insecure instantiations of an ideal-cipher,
the value of the provable security in ideal-cipher model is limited.

\section{Secure Remote Password protocol (SRP)}
\label{sec-srp}
If the underlying group $\group$ in OEKE is indeed a finite field,
then one can instantiate the ideal-cipher with 
$\E_{\beta}(X)=X+\beta$ and obtain the Secure Remote Password protocol 
(SRP6) \cite{srp,srp6}.
But one needs to be careful that SRP protocol uses different values for
the keying material computation which achieves stronger security.
In the SRP6 protocol, the server $\Se$ stores the value
$\langle \C, \beta, s\rangle$ for each client $\C$, where
$\beta=g^v$, $v=\hash(s||\C||\alpha)$, $s$ is a random seed for $\C$,
and $\hash$ is a predetermined hash function.
Assume that the underlying group for the protocol is 
$\group=\Z^*_p=\langle g\rangle$. Then the protocol proceeds as follows:
\begin{enumerate}
\item $\C$ sends his name $\C$ to the server $\Se$.
\item $\Se$ sends $s$ to $\C$.
\item $\C$ chooses a random $x\in [1,ord(g)-1]$ and sends $g^x$ to $\Se$.
\item $\Se$ chooses a random $y\in [1,ord(g)-1]$ and sends $3\beta+g^y$
to $\C$.
\item Let $u=\hash(g^x||3\beta+g^y)$.
$\C$ sends $M=\hash(g^x||3\beta+g^y||S)$ to $\Se$ where $S=g^{y(x+uv)}$.
\item $\Se$ verifies that $M$ is correct and sends $\hash(g^x||M||S)$ to $\C$.
\item $\C$ verifies that $\hash(g^x||M||S)$ is correct.
\item Let $K=\hash(S)$.
\end{enumerate}
The role of $u$ in SRP6 is to defeat an adversary $\A$ who may know $\beta$.
If $\A$ knows $\beta$ and $u$ is fixed, she can impersonate $\C$ by sending
$g^x\cdot g^{-vu}=g^{x-uv}$ instead of $g^x$ in the third step.
Then $g^{y(x-uv+uv)}=g^{xy}$, and $K=\hash(g^{xy})$.
Note that this additional value $u$ in the SRP protocol achieves 
stronger security against stolen $\beta$ while OEKE does not have this
level of security.

If we instantiate the ideal cipher in OEKE with 
$\E_{\beta}(X)=X\cdot \imath(\hash(\beta))$ and use the
SRP6 shared secret computation method, then we get a natural 
generalization of the SRP protocol,
where $\imath$ ``appropriately'' maps a random string to 
a group element of order $ord(g)$. For example, if we define 
$\imath(\hash(\beta))$ by the following procedure, 
then we get the SRP5 protocol \cite{wang1363} which is currently under
standardization in the IEEE 1363.2 standard working group.
\begin{enumerate}
\item Let $x=\hash(\beta)$.
\item\label{stephere} If $x$ is a group element of order $ord(g)$, then let 
$\imath(\hash(\beta))=x$. Otherwise, increase $x$ by one
and go to step (\ref{stephere}). Note that the sentence ``increase $x$ by one''
can be any natural interpretation of ``add one to a group element'' in
a group.
\end{enumerate}
Since the original SRP protocol is based on a 
field and uses both field operations of addition and multiplication, 
there is no direct translation of SRP from
the group $\Z^*_p$ to ECC-based groups. The above
generalization SRP5 of SRP6 can be implemented over ECC groups.

Lee and Lee \cite{ecsrp} have tried to design
ECC-based SRP protocols and introduced four
ECC-based SRP protocols EC-SRP1, EC-SRP2, EC-SRP3, and EC-SRP4.
They used completely different key authentication steps (that is,
the steps (5) to (7) are different). The key steps in their protocols
are the different instantiations of the ideal cipher. That is, they 
recommended replacing the message $3\beta+g^y$ in the fourth step of 
the SRP protocol with the following messages:
\begin{enumerate}
\item $g^y$ for EC-SRP1.
\item $g^\alpha\cdot g^{xy}$ for EC-SRP2.
\item $(g^{x-\alpha})^y$ for EC-SRP3.
\item $(g^{x-\alpha+1})^y$ for EC-SRP4.
\end{enumerate}
The keying material $K$ is the same as that in the original
SRP protocol, i.e.,  $K=\mathrm{SHA}(g^{y(x+uv)})$.
It is straightforward to check that the protocol EC-SRP1 is 
insecure against off-line dictionary attacks.

\section{Conclusions}
\label{conclusions}

In this paper, we presented several examples of real ciphers 
that would result in broken instantiations of the idealized AuthA
and OEKE protocols. Our results show that one should be extremely 
careful when designing or implementing password-based protocols
with provable security in idea-cipher models: a provable security
in ideal-cipher model does not necessarily say that the instantiation
of the protocol is secure.

\section*{Acknowledgements}
The third author would like to thank Prof. Alfred Menezes for
many helpful discussions and comments over the reseach
related to this paper. The authors would also
like to thank the anonymous referees of this journal for
detailed comments on this paper.

\end{document}